\documentclass[10pt,preprint,aps,prc,showpacs]{revtex4-1}
\usepackage{graphicx,epstopdf,multirow,amsmath}
\begin{document}
	  \author{Deepak Kumar}
    \author{Moumita Maiti\footnote{E-mail: moumifph@iitr.ac.in, moumifph@gmail.com (Reprint author)}}
    	\affiliation{Department of Physics, Indian Institute of Technology Roorkree, Roorkee-247667, Uttarakhand, INDIA}
	\author{Susanta Lahiri}
	\affiliation{Chemical Sciences Division, Saha Institute of Nuclear Physics, 1/AF Bidhannagar, Kolkata-700064, INDIA}
		\title{Experimental probe on the production of $^{97}$Ru from $^{7}$Li+$^{93}$Nb reaction:  A study on the precompound emissions}
	\date{\today}
	
	
	\begin{abstract}
		\begin{description}
			\item[Background] 
Interaction of weakly bound heavy-ions with an intermediate or heavy target is not yet understood completely due to the scarcity of experimental data. In order to develop a clear understanding of breakup fusion or preequilibrium emission even at low energy range, 3--10 MeV/nucleon, more experimental investigations are necessary.  
\item[Purpose] 
Study of reaction mechanisms involved in the weakly bound heavy-ion induced reaction, $^7$Li + $^{93}$Nb, at low energies by measuring the production cross sections of the residual radionuclides.
\item[Method] 
Natural niobium ($^{93}$Nb) foil, backed by aluminium (Al) catcher, arranged in a stack was bombarded by $^{7}$Li ions of 20-45 MeV energy.  Activity of the residues produced in each $^{93}$Nb target was measured by off-line $\gamma$-ray spectrometry after the end of bombardment (EOB) and cross sections were calculated. Experimental cross sections were compared with those computed using compound and precompound models. 
\item[Results] 
In general, measured excitation functions of all residues produced in $^{7}$Li + $^{93}$Nb reaction showed good agreement with the model calculations based on Hauser-Feshbach formalism and exciton model for compound and precompound processes, respectively. Significant preequilibrium emission of neutrons was observed at the relatively high energy tail of the excitation function of $^{97}$Ru. 
			
\item[Conclusions] Preequilibrium process played an important role for the enhancement of cross-section in $xn$ reaction channel over the compound reaction mechanism at higher energies for $^{7}$Li + $^{93}$Nb reaction. Additionally, indirect evidence of incomplete or breakup fusion was also perceived.

		 \end{description}
	\end{abstract}
   
  
	\maketitle
     
	
	\section{Introduction}
	\label{L1}
		Study of interaction of weakly bound light heavy-ion induced reactions with intermediate or heavy nuclei at low projectile energies was started about half-century ago. However, complete understanding of the mechanism of heavy-ion reactions is still lacking compared to the light-ion reactions, hence it is a subject of great interest since many years \cite{ref1,ref2,ref3}. Investigation of fusion reactions involving either weakly bound stable nuclei or unstable nuclei far from stability region have  become important to understand the complete fusion (CF) and incomplete fusion (ICF) reactions, nucleon transfer reactions, preequilibrium (PEQ) reactions and quasi-fission because of low nucleon (cluster) separation energies \cite{ref4,ref5,ref6,ref7,ref8}. In addition to that, fusion with weakly bound nuclei is also an important tool to study astrophysical reactions, such as in understanding of nucleosynthesis processes and in studying nuclei near drip line \cite{ref9}. Investigations with weakly bound unstable nuclei are being carried out at radioactive ion beams (RIB) facilities which usually deliver low intense beam. Study of reactions induced by stable nuclei is therefore important, as they produce good quality statistical data, not only to understand reaction dynamics but also for comparison of reaction quantities obtained by weakly bound unstable projectiles. 
		
		Studies of PEQ processes over the compound nuclear reaction is especially important as the particles are emitted prior to statistical equilibrium provide necessary information about the dynamics of the excited composite system and their mechanism to attain statistical equilibrium. Substantial signature of PEQ process has been witnessed in high energy tail of excitation functions of light and heavy-ion induced reactions. However, besides compound and precompound processes, ICF also starts to compete in heavy-ion induced reaction at relatively high energy (10-25 MeV/nucleon) \cite{ref10,ref11,ref12,ref13,ref14}. Birattari et al. \cite{ref10}, Cavinato et al. \cite{ref11}, Vergani et al. \cite{ref12} experimented on the $^{12}$C and/or $^{16}$O induced reactions on different targets and observed PEQ emission of nucleons during the thermalization of compound system. Moreover, PEQ process was also observed at relatively low energy $\sim$ 4-8 MeV/nucleon where pure evaporation process is dominant. PEQ emission of $\alpha$-particles was reported by Amorini et al. \cite{ref15} in complete and incomplete fusion reaction in the $^{12}$C + $^{64}$Ni reaction at 8 MeV/nucleon. Sharma et al. \cite{ref16} analyzed the PEQ emission of neutrons from $^{12}$C and $^{16}$O induced reaction on $^{128}$Te, $^{169}$Tm, $^{159}$Tb and $^{181}$Ta targets at 4--7 MeV/nucleon. Therefore more experimental investigation near the barrier is necessary to draw specific conclusion and to develop sophisticated theory for PEQ and CF-ICF processes in weakly bound nuclear reactions.
		
		Among the ruthenium isotopes, neutron deficient $^{97}$Ru has the potential, owing to its low lying intense gamma lines: 215.70 keV (85.62\%) and 324.49 keV (10.79\%) energy and moderate half-life (2.83 d), to be used in several applications. Ability of forming wide varieties of chemical complexes made $^{97}$Ru lucrative to the nuclear medicine community. Moreover it can be produced in the no-carrier-added state which is the prerequisite of such applications \cite{ref17}.  
		
 		Its production by neutron or light-ion induced reactions (like $p$, $\alpha$, $^{3}$He) were investigated earlier by several groups \cite{ref18,ref19,ref20,ref21,ref22,ref23}.  $^{97}$Ru was prepared from the high energy proton spallation (200 MeV or 67.5 MeV ) on natural rhodium ($^{103}$Rh) target through $^{103}$Rh($p$, 2$p$5$n$)$^{97}$Ru reaction \cite{ref18,ref19} along with the radionuclides of Tc, Rh and Pd as impurity. Enormous production of $^{97}$Ru was reported from 50 MeV proton induced reaction on radioactive target $^{99}$Tc \cite{ref20}. Besides proton, $\alpha$-particle or $^3$He induced reactions on natural molybdenum target also led to the production of $^{97}$Ru along with Tc and Ru contaminants via $^{nat}$Mo($^{4}$He, $xn$)$^{97}$Ru, and  $^{nat}$Mo($^{3}$He, $xn$)$^{97}$Ru reactions, respectively, \cite{ref21,ref22,ref23}. Although enriched $^{96}$Ru is expensive, the most easiest way to produce $^{97}$Ru is by thermal neutron capture reaction, $^{96}$Ru($n$,$\gamma$)$^{97}$Ru, but, this leads to the low specific activity of $^{97}$Ru.  

Recently, heavy-ion ($^{7}$Li, $^{12}$C) induced productions of $^{97}$Ru on natural Nb and Y were investigated by Maiti et al. and the subsequent chemical separation of $^{97}$Ru from the target matrix was developed \cite{ref24,ref25,ref26}.  In this article, we have made an effort to study (i) the relevance of PEQ / CF-ICF mechanism in a light heavy-ion induced reaction, $^{7}$Li + $^{93}$Nb, at low energy range 20-42 MeV, and (ii) the production of $^{97}$Ru along with the coproduced radionuclides at various impinging energies, which is essential to determine the optimized production parameters for $^{97}$Ru.				

The experimental procedure and brief of the nuclear model calculations are presented in Sec. \ref{L2} and \ref{L3}, respectively. Section \ref{L4} discusses the results of the present study and Sec. \ref{L5} concludes the report.

	\section{Experimental}
		\label{L2}
		\subsection{Measurement of Activity}
    The $^{7}$Li-ion beam up to 45 MeV energy obtained from BARC-TIFR Pelletron Accelerator facility, Mumbai, India, was used for the experiment. Spectroscopically pure (99.99\%) natural niobium ($^{93}$Nb) was procured from Alfa Aesar and self supporting Nb-foils of 2.3--3.2 mg/cm$^{2}$ were prepared by the proper rolling. The niobium and aluminum ($^{27}$Al) foils were mounted on an aluminum ring of 12 mm inner and 22 mm outer diameter with 0.5 mm thickness. The $^{7}$Li$^{3+}$--ion beam was allowed to incident on niobium targets backed by Al foils of $\sim$1.5 mg/cm$^{2}$ arranged in a stack. A total of six such Nb-Al foils stack was irradiated individually varying the incident energy of $^{7}$Li$^{3+}$--ions with a slight overlap between them. Total charge of each irradiation was measured by an electron-suppressed Faraday cup placed at the rear of the target assembly. Use of Al foil served the purpose of an energy degrader as well as catcher for recoils, if any, in the beam direction. The large area of the catcher foils ensured the complete collection of recoiled evaporation residues.  The duration of the irradiation time was chosen according to the beam intensity and half-lives of the product radionuclides. Energy degradation in each foil was estimated by Stopping and Range of Ions in Matter (SRIM) code \cite{ref27}. The projectile energy at a target is estimated by averaging the incident and outgoing beam energy.  
	  
	After the end of bombardment (EOB), target $^{93}$Nb and catcher $^{27}$Al foils were assayed using off-line $\gamma$-spectrometry in a regular time interval for a sufficient time to measure the activity of the residues with the help of Falcon 5000, BEGe-based detector, having enhanced efficiency and resolution at low energy while still preserving good efficiencies at high energies, coupled with a PC operating with GENIE-2K software (Canberra).  Detector was calibrated using the standard sources, $^{152}$Eu (13.506 a), $^{137}$Cs (30.08 a), $^{60}$Co (5.27 a), $^{133}$Ba (10.51 a), of known activity. Energy resolution of the detector was $\le$ 2.0 keV at 1332 keV energy.
	Background subtracted peak area count corresponding to a particular $\gamma$-ray energy is the measure of yield of an evaporation residue \cite{ref28}. 
	
	The cross section of the $\textit{n}^{th}$ evaporation residue, $\sigma_{\textit{n}}$(\textit{E}), at an incident energy, \textit{E} is calculated from the equation
\begin{equation}
\label{mmeq2}
\sigma_{n}(E)=\frac{Y_{n}}{I_{pro}N_{tg}x_{tg} (1-e^{-\lambda_{n}T})}
\end{equation} 
The yield (\textit{Y$_{n}$}) of an evaporation residue $\textit{n}$ at the EOB was calculated from the equation

\begin{equation}
\label{mmeq1}
Y_{n}=\frac{C(t)}{\varepsilon^{\gamma}_{n}I^{\gamma}_{n}}e^{\lambda_{n} \tau} 
\end{equation} 
where $\textit{C}(\textit{t})$ is the count rate (count per second), $\varepsilon^{\textit{n}}_{\gamma}$ and $\textit{I}^{\textit{n}}_{\gamma}$ are the detection efficiency and branching intensity of the characteristic $\gamma$--ray of the evaporation residue, decay constant is $\lambda^{\textit{n}}$, cooling time is $\tau$. 
$I_{pro}$ is the beam intensity of the projectile ions, $N_{tg}$ and $x_{tg}$ are the number of target nuclei per unit volume and target thickness, respectively, $T$ is the duration of irradiation \cite{ref29,ref30}. 
The nuclear spectroscopic data used to calculate the production cross sections of the evaporation residue are enlisted in the Table \ref{t1} \cite{ref31}.

\subsection{Estimation of uncertainties}
	Uncertainties in the cross section measurement may come from the following: (i) inaccuracy in efficiency calibration of the detector $\sim$ 2\%, (ii) non-uniformity of samples and measurement of its thickness in atoms/cm$^{2}$ may cause error $\sim$ 5\%, (iii) uncertainty in the beam current measurement was $\sim$ 5\% (iv) error propagated to the cross section measurement from the counting statistics, which is negligible in this case, (v) error in the estimation of beam energy due to the degradation of energy while traversing through the successive target foils, however, energy straggling effects is expected to be very small and are neglected in the calculation \cite{ref32, ref33}. The total uncertainty associated in to cross section measurement was determined considering all those factors and the data presented upto 95\% confidence level.

	\section{Theoretical calculation}
	\label{L3}
	
	Nuclear reaction can be broadly classified into three reaction mechanisms: direct (\textsc{DIR}), preequilibrium (\textsc{PEQ}) and equilibrium or evaporation (\textsc{EQ}). Production of a residual nucleus in a nuclear reaction is the contribution from all three types. In this endeavor, an effort has been made to explain the measured cross-section data of the residues produced in the $^{7}$Li + $^{93}$Nb reaction in terms of \textsc{PEQ} and \textsc{EQ} reactions in the 20--45 MeV energy range using nuclear reaction model codes \textsc{PACE4} \cite{ref34} and \textsc{ALICE91} \cite{ref35, ref36}, and \textsc{ EMPIRE3.2} \cite{ref37}. In general, contribution of \textsc{DIR} reaction is not expected at low incident energies. 

		\subsubsection{PACE4} 
    \textsc{PACE4}, is based on Hauser-Feshbach formalism which follows the correct procedure of angular momentum coupling at each stage of deexcitation of an excited nuclei. For heavy projectile, fusion cross section and initial spin distribution is calculated by Bass model \cite{ref38} while optical model is used for light ions. However heavy-ion fusion near and below the barrier and reaction induced by very heavy beams can not be determined by Bass Model. 
		The transmission coefficients for light particle emission are created by the optical model calculations where all the optical model parameters are taken from Ref. \cite{ref39}. The shift in the coulomb barrier during deexcitation is accounted by calculating the transmission coefficients at an effective energy determined by the shift. Fission is considered as a decay mode, and the fission barrier can be changed accordingly in the program.  The Gilbert-Cameron level density is used in the calculation, with level density parameter, $ a = A/10 $, where $A$ is mass number of compound nucleus. Little $a$ ratio, $a_{f}$/$a_{n}$, is taken as unity. A. J. Seirk modified rotating liquid drop barrier is adopted. A non-statistical yrast cascade gamma decay chain is artificially incorporated to simulate gamma multiplicity.
	   
	\subsubsection{ALICE91} 
	\textsc{ALICE91} has been used to study EQ and PEQ emission of particles in the $^{7}$Li + $^{93}$Nb reaction. Hybrid or geometry dependent hybrid model \cite{ref40} computes the PEQ emission of particles and Weisskopf-Ewing model \cite{ref41} accounts the compound emission process. It does not account for the direct reaction processes. In hybrid model, emission of particles results from the two body interaction process in an excited projectile-target composite system. Each stage of the relaxation process is specified by the exciton number ($n_0$) of excited particles, i.e., sum of excited particles ($p$) and holes ($h$). The hybrid model uses “never-come-back” approximation, i.e., in each two-body interaction, $p$-$h$ pairs may either be created or redistribution of energy may take place among the excitons. It explicitly determines the PEQ emission energy distribution of the excited particles, which helps to estimate high energy emissions more accurately.  Details of hybrid model is available elsewhere \cite {ref17, ref28}. Geometry dependent hybrid model is selected for the calculation to include the nuclear surface effect. In \textsc{ALICE91}, light particles emission ($n$, $p$, $d$ etc.) from equilibrated nucleus are calculated upto 12 mass units wide and 10 charge units deep from the composite nucleus system. Fermi gas level density is used for the cross-section calculation with level density parameter, $a$ = $A$/9 MeV$^{-1}$. Optical model is used for the calculation of inverse reaction cross section. The rotating finite-range fission barriers of Sierk have been selected. The total number of nucleons in the projectile has been chosen as the initial exciton number for the PEQ cross section calculation.
	   
    \subsubsection{EMPIRE3.2} 
		\textsc{EMPIRE3.2} code accounts all the three major nuclear reactions -- EQ, PEQ and DIR. For compound reaction process, detail Hauser-Feshbach model, which follow the exact coupling of angular momentum and parity of emitted particles and residual nucleus, is used including width fluctuations and the optical model for fission. PEQ emission can be calculated either by quantum mechanical PEQ models (multi-step direct (MSD) or multi-step compound (MSC) mechanism \cite{ref42}) or by phenomenological PEQ models (exciton model or hybrid Monte Carlo simulation \cite{ref43}). Coupled channels approach or distorted wave born approximation (DWBA) \cite{ref44, ref45} is used for the calculation of direct processes. The code can be applied to the calculation of neutron capture in the keV region, as well as for heavy-ion induced reactions at several hundreds of MeV. Coupled-Channels calculation (\textsc{CCFUS}) \cite{ref46} is used for heavy ion fusion cross section. Nuclear masses, optical model parameters, ground state deformations, discrete levels and decay schemes, level densities, fission barriers, and $\gamma$-ray strength functions are internally provided by input library \textsc{RIPL-3}. In our calculation, exciton model is used for PEQ emission process and enhanced generalized superfluid model level density (\textsc{EGSM}) is used to consider the collective (rotational/vibrational) effect of nuclei on nuclear level density.
		
		In the EGSM, effect of superconducting pairing correlations, which strongly influence the nuclear level density at lower energy, is considered as a correlation function $\delta_0$.\ The EGSM is build on Fermi Gas Model (FGM) level density in an adiabatic mode along with collective enhancement factor which damp out with increasing excitation energy ($E_x$) and reduces to unity above critical temperature (T$_c$), that is, it reduces naturally to FGM above T$_c$. In this model, critical level density parameter ($a_c$) is used below T$_c$, while Ignatyuk empirical level density parameter, 
	$a$($E_x$)= $\tilde{a}$ [ 1 + (1 - e$^{-\gamma_s U^*}$){$\delta S$}/{$U^*$}]
	is used above T$_c$, where parameters $\tilde{a} = 0.0748A$ and $\gamma_s = 0.5609 A^{1/3}$ are the asymptotic value of $a$-parameter and shell effects damping parameter, respectively. $\delta{S}$ is the shell correction which fades out with increasing excitation energy ($E_x$) and $U^* = U - {0.1521}{a_c} {\delta_0^2}$, is the effective energy above T$_c$, while below T$_c$, U is used as effective energy, $U = E_x + n{\delta_0}$, where correlation function is calculated as $\delta_0 = 12/\sqrt{A}$, and n=0, 1 and 2 for odd-odd, odd-A and even-even nuclei, respectively.
		 
	\section{Discussion on results}
	\label{L4}
\begin{figure} 
\includegraphics [width=10.6cm]{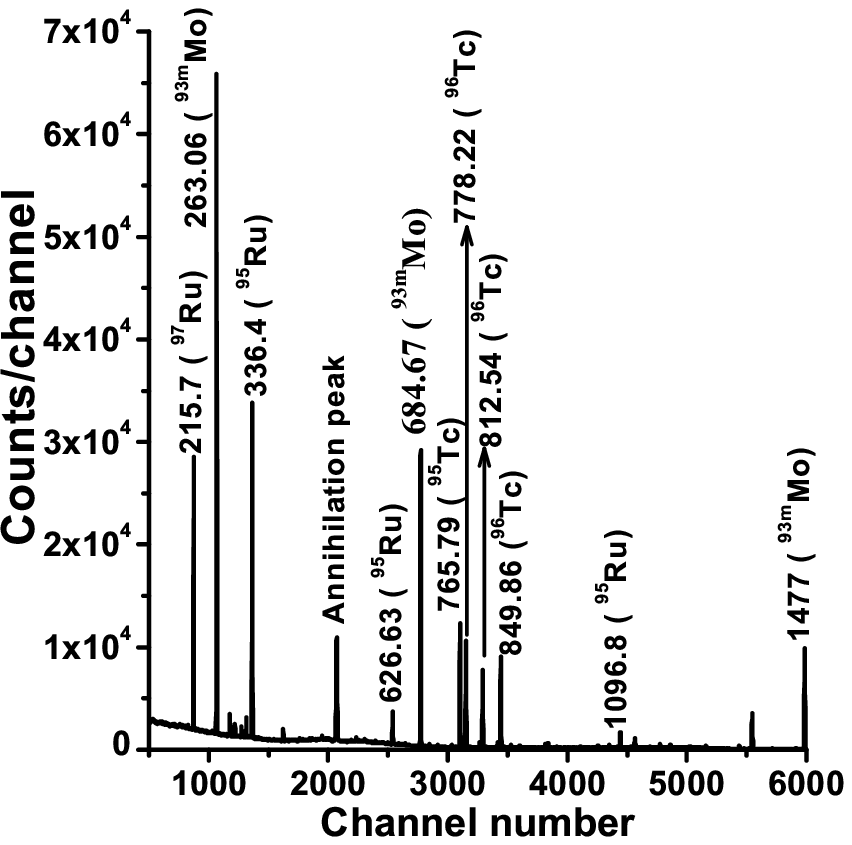}
\caption{ $\gamma$-ray spectrum of $^{7}$Li activated niobium foil collected 34 minutes after the EOB }
\label{fig1}
\end{figure}

\begin{figure}
\includegraphics [width=8.6cm]{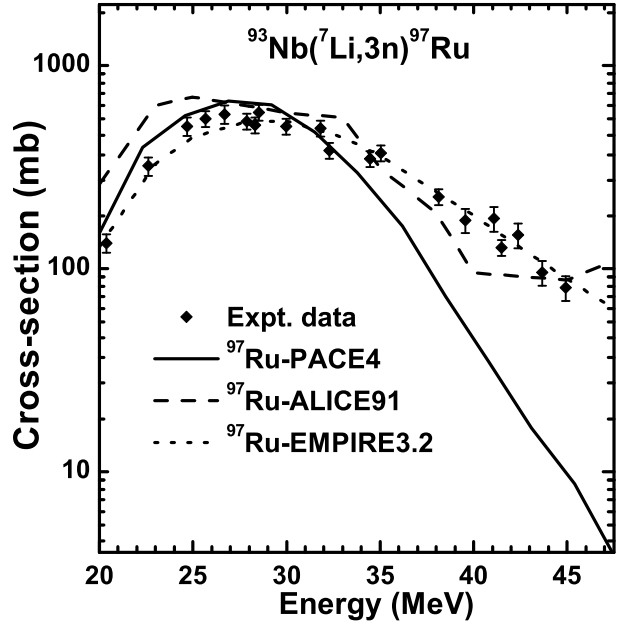}
\caption{\label{fig:}Comparison of experimental (symbol) excitation functions of $^{97}$Ru from $^{7}$Li + $^{93}$Nb reaction and those obtained from theoretical (curves) estimation from \textsc{PACE4}, \textsc{ALICE91} , and \textsc{EMPIRE3.2}}
\label{fig2}
\end{figure}

\begin{figure}
\includegraphics[width=8.6cm]{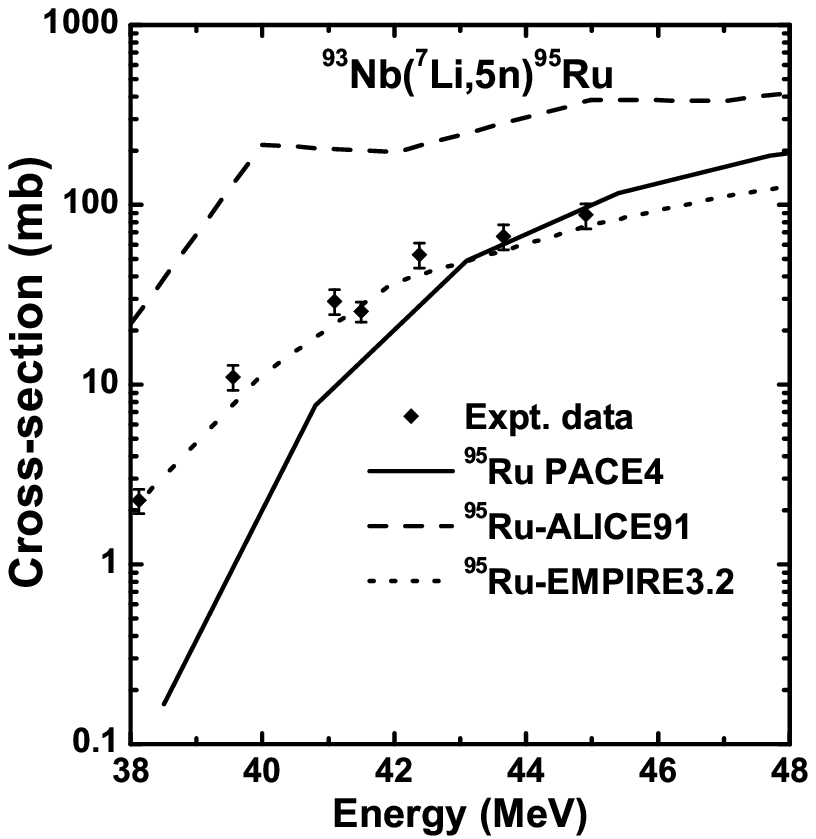}
\caption{Comparison of experimental (symbol) and calculated (curves) excitation functions for production of $^{95}$Ru}
\label{fig3}
\end{figure}

\begin{figure} 
\includegraphics[width=8.6cm]{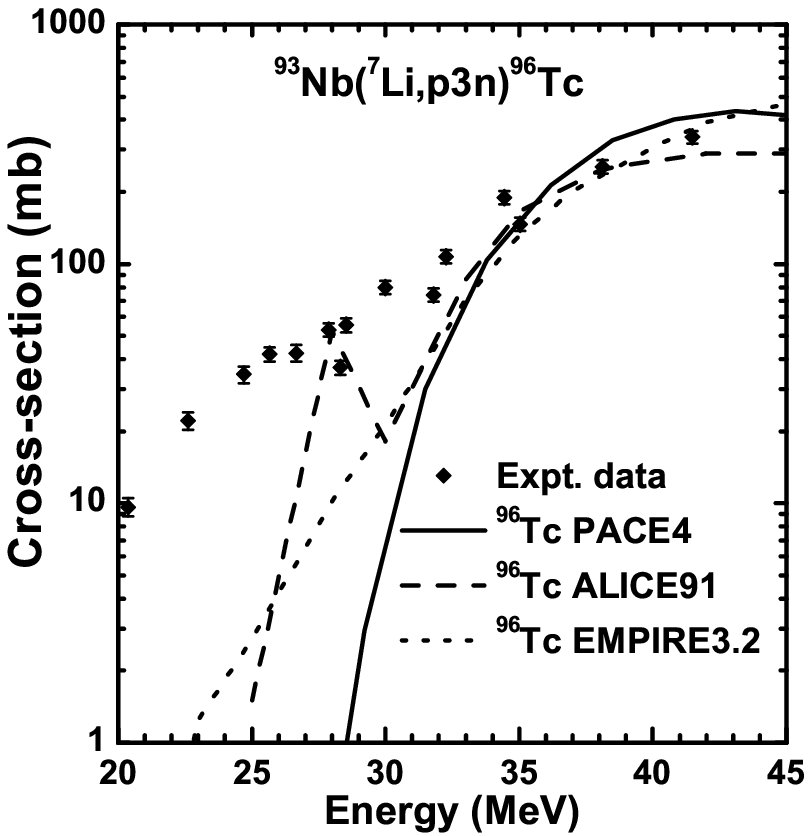}
\caption{Comparison of experimental (symbol) and calculated (curves) excitation functions for production of $^{96}$Tc}
\label{fig4}
\end{figure}

\begin{figure} 
\includegraphics [width=8.6cm]{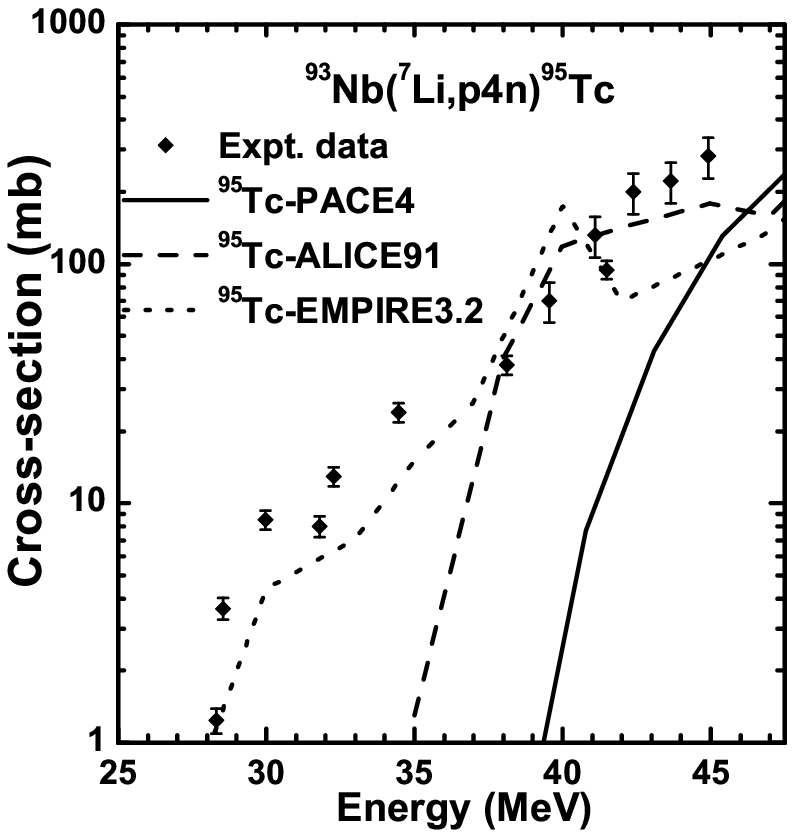}
\caption{Comparison of experimental (symbol) and calculated (curves) excitation functions for production of $^{95}$Tc}
\label{fig5}
\end{figure}

\begin{figure}
\includegraphics [width=8.6cm]{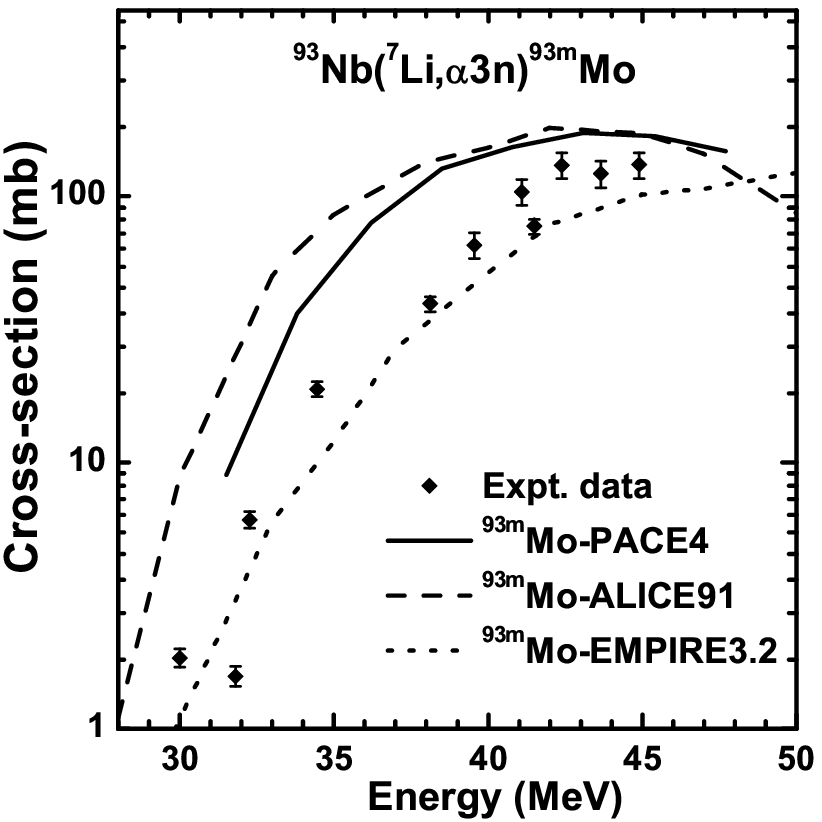}
\caption{Comparison of experimental (symbol) and calculated (curves) excitation functions for production of $^{93m}$Mo}
\label{fig6}
\end{figure}

	Analysis of the time resolved $\gamma$--ray spectra collected after EOB was carried out for each set of Nb-Al foils to identify the residual radionuclides  produced in the $^{7}$Li + $^{93}$Nb reaction at different incident energies. It ensured the production of  $^{97}$Ru, $^{95}$Ru, $^{96}$Tc, $^{95}$Tc, and $^{93m}$Mo in the target matrix.
		A typical $\gamma$--ray spectrum of the evaporation residues produced in the $^{7}$Li + $^{93}$Nb reaction at $42$ MeV  incident energy collected 34 minutes after the EOB is presented in Fig \ref{fig1} with their characteristics $\gamma$--rays. The possible reactions contributing to the production of the residues are listed in Table \ref{t1} along with the reaction threshold. Measured cross sections of the evaporation residues at various energies are listed in Table \ref{t2}. Comparison between the experimental excitation functions of the residues and those theoretically computed using the nuclear reaction model codes \textsc{PACE4} \cite{ref34} and \textsc{ALICE91} \cite{ref35,ref36}, and \textsc{EMPIRE3.2} \cite{ref37} are shown in Figs \ref{fig2}--\ref{fig6}. Experimentally measured cross sections are shown by symbol with the error bar, while theoretical calculations  are shown by curves. 
	
    Figure \ref{fig2} shows the production cross sections of $^{97}$Ru from 20--45 MeV energy range. It is observed that, at low energies, experimental cross sections are well reproduced by all three theoretical calculations. However, at higher energy region ($\sim$5--7 MeV/nucleon) a clear deviation is observed between the measured cross sections and \textsc{PACE4} estimation, while \textsc{ALICE91} and \textsc{EMPIRE3.2} are in good agreement with the experimental data. The reason is that \textsc{PACE4} computation is based only on the compound nuclear model using Hauser-Feshbach formalism, whereas \textsc{ALICE91} and \textsc{EMPIRE3.2} both considered PEQ as well as compound nuclear model in the calculation.  It is evident that significant PEQ emission occurs around the 5--7 MeV/nucleon energy region. A critical observation also shows that \textsc{EMPIRE3.2} prediction reproduced the experimental cross section more accurately  than the \textsc{ALICE91}.
		
Comparison of measured and theoretical excitation functions of $^{95}$Ru is shown in Fig \ref{fig3}. Experimental data agree well with \textsc{EMPIRE3.2} calculation throughout the measured energy range but \textsc{PACE4} underpredicts the measured data below 42 MeV. This might be due to inclusion of enhanced generalized super-fluid model density in \textsc{EMPIRE3.2} as it accounts the collective (rotational/vibrational) effect of the nuclear level density which enhance the nuclear level density below the critical energy. \textsc{ALICE91} overpredicts the data about four times over the energy range studied. 
The PEQ emission is observed in the 3$n$ reaction channel (Fig \ref{fig2}), unlike 5$n$ reaction channel. It is anticipated that one PEQ neutron emission is more likely than two or more near the barrier energy, hence PEQ emission of one neutron from an excited composite nuclear system is possible even at low projectile energy.  
	 
   Figure \ref{fig4} represents the excitation function of $^{96}$Tc radionuclides in 20--45 MeV energy interval. All three theoretical estimations reproduce the experimental data at higher energy region, but underpredicts the cross sections at low energies. Besides complete fusion and PEQ mechanism , the higher experimental cross sections of $^{96}$Tc at the low energy region might be attributed to the incomplete fusion (ICF) process, which is likely to occur in the interaction of weakly bound projectile $^{7}$Li with $^{93}$Nb.
Thus $^{96}$Tc might be produced by following possible reaction channels -- \\
	 1. Complete fusion of $^{7}$Li with $^{93}$Nb leads to production of the $^{96}$Tc through $p$3$n$ channel
	         	\begin{equation}
	         	  \begin{aligned}
	         	^7Li+^{93}Nb \rightarrow [^{100}Ru] \rightarrow ^{96}Tc + p3n,\\
	      	         E_{th} = 19.3 \ MeV.
	      	      \end{aligned}   
	         	\end{equation}      	
	 2. Complete fusion of $^{7}$Li with $^{93}$Nb leads to production of the $^{96}$Tc by $d$2$n$ channel
	         	\begin{equation}
	         	   \begin{aligned}
	         		^{7}Li+^{93}Nb \rightarrow [^{100}Ru] \rightarrow ^{96}Tc + d2n,\\
	        	         		E_{th} = 16.9 \ MeV. 
	               \end{aligned}  
	         	\end{equation}    	
	 3. Complete fusion of $^{7}$Li with $^{93}$Nb leads to production of the $^{96}$Tc by $tn$ channel
	 	 \begin{equation}
	 	 \begin{aligned}
	 	 ^{7}Li+^{93}Nb \rightarrow [^{100}Ru] \rightarrow ^{96}Tc + tn,\\
	 	 E_{th} = 10.2 \ MeV. 
	 	 \end{aligned}  
	 	 \end{equation}        	       	
	 4. It is possible that $^{7}$Li dissociates into $\alpha$-particle and tritium in the nuclear force field.  $\alpha$-particle, the secondary projectile, fuses with $^{93}$Nb forming a composite nucleus  $^{97}$Tc$^*$,which emits one neutron to form $^{96}$Tc, and tritium  moves in the forward direction as a spectator.
	         \begin{equation}
	        	\begin{aligned}
	        		^7Li(^4He + t) \rightarrow ^4He + ^{93}Nb  \rightarrow [^{97}Tc^*] \rightarrow ^{96}Tc + n,\\
	       	        		E_{th} = 7.3\ MeV. 
	       	    \end{aligned} 
             \end{equation}            
	5. Interaction of $^{7}$Li with $^{93}$Nb may also lead to the production of $t$ and $^{97}$Tc in the excited level, which may emit one neutron to produce $^{96}$Tc.
	 	 \begin{equation}
	 	 \begin{aligned}
	 	 ^{7}Li+^{93}Nb \rightarrow t+^{97}Tc (E_{th}=39.6\ MeV) \rightarrow ^{96}Tc + n,\\
	 	 \end{aligned}  
	 	 \end{equation}        	
	 
	Excitation function for $^{95}$Tc residue is plotted in Fig \ref{fig5}. \textsc{PACE4} calculations underpredict the experimental excitation function throughout the range. Although \textsc{ALICE91} explains measured data at higher energy region but underpredicts at lower energy region. However, \textsc{EMPIRE3.2} calculations show a good agreement to the experimental data even at lower energies. 	 
	Figure \ref{fig6} shows the production of $^{93m}$Mo radionuclide. \textsc{PACE4} and \textsc{ALICE91} overpredict the experimental data throughout the energy region, while \textsc{EMPIRE3.2} reproduced the experimental data successfully. 
	
	It is remarkable that \textsc{EMPIRE3.2} calculation are in good agreement with the measured excitation functions of all the residues. It projects the effectiveness of \textsc{EMPIRE3.2} nuclear reaction code in understanding the heavy-ion induced reaction in the low and intermediate energy range. \textsc{ALICE91} was intended only to study light-ion ($n$, $p$, $d$, $\alpha$-particle) induced reactions while \textsc{EMPIRE3.2} code is competent for both light as well as heavy ion induced reactions.

     From the measured excitation functions, it is seen that production of neutron deficient $^{97}$Ru radionuclides between 22--35 MeV energy range is high compared to other coproduced radionuclides $^{96}$Tc and $^{95}$Tc, which along with bulk Nb can be chemically separated easily from the $^{97}$Ru \cite{ref24}. Maximum cross section ($\sim$ 580 mb) of $^{97}$Ru was observed at 28.5 MeV energy along with one tenth of $^{96}$Tc radioisotopes.

   	\section{Conclusion}
		\label{L5}
	Production cross sections of all the residual radionuclides produced in the $^{7}$Li + $^{93}$Nb reaction have been studied in the 20--45 MeV energy range and are compared with the theoretical model calculations -- \textsc{PACE4}, \textsc{ALICE91} and \textsc{EMPIRE3.2} with the suitable choice of parameters. Overall, \textsc{EMPIRE3.2} estimations agree well with all the measured excitation functions. Measured cross section data indicate the compound nuclear reaction process as a predominant mechanism. However, significant PEQ emission of neutrons was also observed in the high energy tail of excitation function in the 3$n$ emission channel. Therefore, higher values of cross section data at the high energy tail could only be explained by the contributions of compound and PEQ process. In order to understand the PEQ emission in $xn$, $x\ge$ 5 channel, experimental data is needed at higher energy region. Further, indirect signature of incomplete fusion was also observed in the production of $^{96}$Tc radioisotope. Since $^{7}$Li is a weakly bound projectile and it can easily break into $\alpha$-particle and tritium. It is expected that breakup fusion of $\alpha$-particle might have taken place with $^{93}$Nb and the subsequent emission of neutron by compound or PEQ process may produce $^{96}$Tc. However proper investigation of breakup fusion such as recoil range distribution method is needed for the confirmation of incomplete fusion in $^{7}$Li + $^{93}$Nb system.


	\begin{acknowledgments}
		Authors convey their sincere thanks to the Pelletron staff and target laboratory staff of BARC-TIFR Pelletron facility for their cooperation and help during the experiment. The work is financially supported by MHRD, Government of India, IITR/SRIC/218/FIG, SINP-DAE-12-plan project TULIP grant.  
	\end{acknowledgments}


\begin{table*}
\caption{Spectroscopic data \cite{ref31} of the residual radionuclides and list of contributing reactions}
\label{t1}
\begin{ruledtabular}
\begin{center}
					\renewcommand{\arraystretch}{1.2}
					\begin{tabular}{cccccc}
						Nuclides(J$^\pi$) & Half-life & Decay mode (\%)  & E$\gamma$(keV)[I$\gamma$(\%)] & Reactions & E$_{th}$(MeV)~\footnote{E$_{th}$ represents threshold energy.}\\
						\hline
						$^{97}$Ru(5/2$^+$) & 2.83 d & $\epsilon$(100) &  215.7[85.6] & $^{93}$Nb($^{7}$Li, 3n) & 11.2\\
						&    &    & 324.5[10.8] \\		
						$^{95}$Ru(5/2$^+$) & 1.64 h & $\epsilon$(100) & 336.4[70.2] & $^{93}$Nb($^{7}$Li, 5n)& 31.4\\
						&     &    & 626.6[17.8] & &\\
						$^{96}$Tc(7$^+$) & 4.28 d & $\epsilon$(100)  & 778.2 [99.8] & $^{93}$Nb($^{7}$Li, p3n)& 19.3\\
						&    &   & 812.5[82] & $^{93}$Nb($^{7}$Li, d2n)           & 16.9\\
						&    &   & 849.9[98] &$^{93}$Nb($^{7}$Li, tn)  &   10.2\\
						&		 &   &           & $^{93}$Nb($^{7}$Li, t)$^{97}$Tc$\rightarrow$$^{96}$Tc+n &   39.6\\
						&    &   & 					 &$^{93}$Nb($\alpha$, n) - ICF  &   7.3\\
					 $^{95}$Tc(9/2$^+$) & 20 h & $\epsilon$(100) &  765.8[93.8] & $^{93}$Nb($^{7}$Li, p4n) & 27.8\\
						&  &  &  &$^{93}$Nb($^{7}$Li, d2n) &   25.4 \\
						$^{93m}$Mo(21/2$^+$) & 6.85 h& IT(99.88)$\epsilon$(0.12)  & 263.1[56.7] & $^{93}$Nb($^{7}$Li, $\alpha$3n) & 13.1\\
						&     &   & 684.7[99.7] & $^{93}$Nb($^{7}$Li, 2p5n) &   43.5\\
					\end{tabular}
				\end{center}
			\end{ruledtabular}
		\end{table*}
		
		\begin{table*}{}
	   \caption{Cross-section (mb) of residues at different incident energies}
	    \label{t2}
	    	\begin{ruledtabular}
	    		\begin{center}
						\renewcommand{\arraystretch}{1.2}
	    			\begin{tabular}{cccccc}
	    				\multirow{3}{1.5cm}{\bfseries Energy (MeV)} \\
	    				& \multicolumn{5}{c} { \bfseries Cross-section (mb)}\\
	    				\cline{2-6}
	    				& \multicolumn{1}{c}{$^{97}$Ru}
	    				& \multicolumn{1}{c}{$^{95}$Ru}
	    				& \multicolumn{1}{c}{$^{96}$Tc}
	    				& \multicolumn{1}{c}{$^{95}$Tc}
	    				& \multicolumn{1}{c}{$^{93m}$Mo}\\			\hline
	    				20.4 & 133.6 $\pm$ 14 &                  & 9.7 $\pm$ 0.8    &  & \\
	    				22.6 & 320 $\pm$ 33.3 &                  & 22.1 $\pm$ 1.8   & & \\
	    				24.7 & 503.3 $\pm$ 52.1 &                  & 34.6  $\pm$ 2.7  & &  \\
	    				25.7 & 547.3 $\pm$ 48.9 &                  & 42 $\pm$  2.9  & & \\	
	    				26.7 & 575.3 $\pm$ 59.6 &                  &  42.5 $\pm$ 3.3  & & \\
	    				27.9 & 530.3 $\pm$ 47.4 &                  & 53.1 $\pm$ 3.4   &  &  \\
	    				28.3 & 508.4 $\pm$ 45.5 &                  & 36.9 $\pm$ 2.5   & 1.2 $\pm$ 0.2 &  \\
	    				28.5 & 587.3 $\pm$ 52.7 &                  & 55.6 $\pm$ 3.7   & 3.6 $\pm$ 0.4 & 0.2 $\pm$ 0.1 \\
	    				30.0 & 502.2 $\pm$ 44.9 &                  & 79.8 $\pm$  5.1  & 8.5 $\pm$ 0.8 & 1.8 $\pm$ 0.2 \\
	    				31.8 & 492.6 $\pm$ 44.1 &                  & 74.3 $\pm$ 4.8   & 8 $\pm$ 0.8 & 1.6 $\pm$ 0.1  \\
	    				32.3 & 382.3 $\pm$ 34.2 &                  &  107.5 $\pm$ 6.8 & 13 $\pm$ 1.2  & 6.1 $\pm$ 0.4  \\	
	    				35.1 & 371.2 $\pm$ 33.4 &                  & 146.4 $\pm$ 9.3  & 19.4 $\pm$ 1.8 & 12.3 $\pm$ 0.9 \\
	    				38.1 & 226.3 $\pm$ 20.4 &  2.3 $\pm$ 0.4 & 254.3 $\pm$ 16 & 37.8 $\pm$ 3.4 & 39.5 $\pm$ 2.6 \\
	    				39.5 & 173.4 $\pm$ 24 & 11 $\pm$ 1.8 & 339.9 $\pm$ 37.8 & 70.3 $\pm$ 13.5 & 65.6 $\pm$ 7.3 \\	
	    				41.1 & 176.9 $\pm$ 24.5 & 29 $\pm$ 4.6 & 478.9 $\pm$ 53.2 & 131.5 $\pm$ 25.3 &104.1 $\pm$ 11.6 \\
	    				41.5 & 126.6 $\pm$ 11.5 & 25.6 $\pm$ 3.3 & 338.9 $\pm$ 21.2 & 94.6 $\pm$ 18.4 & 77.2 $\pm$ 4.9 \\	
	    				42.4 & 146.2 $\pm$ 24.5 & 53 $\pm$ 8.4 & 522.9 $\pm$ 58.2 & 199.4 $\pm$ 38.4 & 131.4 $\pm$ 14.7 \\
	    				43.7 & 95.5 $\pm$ 13.2  & 67 $\pm$ 10.6& 424.3 $\pm$ 47.3 & 222.3 $\pm$ 42.8 & 121.8 $\pm$ 13.6 \\	
	    				44.9 & 80.6 $\pm$ 11.2  & 87.5 $\pm$ 13.9& 413.5 $\pm$ 46.1 & 281.8 $\pm$ 54.3 & 131.7 $\pm$ 14.7 \\
	    			\end{tabular}
	    		\end{center}
	    	\end{ruledtabular}
	    \end{table*}


\begin{thebibliography}{9}
	 \bibitem{ref1} H. C. Britt, A. R. Quinton, Phys. Rev. 124, 877 (1961).
	 \bibitem{ref2} D. J. Parker, J. Asher, T. W. Conlon, I. Naqib, Phys. Rev. C 30, 143 (1984).
	 \bibitem{ref3} D. J. Parker, J. J. Hogan, J. Asher, Phys. Rev. C 39, 2256 (1989).
	 \bibitem{ref4} M. Dasgupta, D. J. Hinde, N. Rowley, A. M. Stefanini, Ann. Rev. Nucl. Part. Sci. 48, 401 (1998).
	 \bibitem{ref5} J. F. Liang, C. Signorini, Int. J. Mod. Phys. E 14, 1121 (2005).
	 \bibitem{ref6} L. F. Canto, P. R. S. Gomes, R. Donangelo, M. S. Hussein, Phys. Rep. 424, 1 (2006).
	 \bibitem{ref7} N. Keeley, R. Raabe, N. Alamanos, J. L. Sida, Prog. Part. Nucl. Phys. 59, 579 (2007).
	 \bibitem{ref8} L. F. Canto, P. R. S. Gomes, R. Donangelo, J. Lubian, M. S. Hussein, Phys. Rep. 596, 1 (2015).
	 \bibitem{ref9} K. E. Rehm, H. Esbensen, C. L. Jiang, B. B. Back, F. Borasi, B. Harss, R. V. F. Janssens, V. Nanal, J. Nolen, R. C. Pardo, M. Paul, P. Reiter, R. E. Segal, A. Sonzogni, J. Uusitalo, A. H. Wousmaa, Phys. Rev. Lett. 81, 3341 (1998).
	 \bibitem{ref10} C. Birattari, M. Bonardi, M. Cavinato, E. Fabrici, E. Gadioli, E. Gadioli Erba, F. Groppi, M. Bello, C. Bovati, A. Di Filippo, T. G. Stevens, S. H. Connell, J. P. F. Sellschop, S. J. Mills, F. M. Nortier, G. F. Steyn, C. Marchetta, Phys. Rev. C 54, 3051 (1996).
	 \bibitem{ref11} M. Cavinato, E. Fabrici, E. Gadioli, E. Gadioli Erba, P. Vergani, M. Crippa, G. Colombo, I. Redaelli, M. Ripamonti, Phys. Rev. C 52, 2577 (1995).
	 \bibitem{ref12} P. Vergani, E. Gadioli, E. Vaciago, E. Fabrici, E. Gadioli Erba, M. Galmarini, G. Ciavola, C. Marchetta, Phys. Rev. C 48, 1815 (1993).
	 \bibitem{ref13} M. Dasgupta, P. R. S. Gomes, D. J. Hinde, S. B. Moraes, R. M. Anjos, A. C. Berriman, R. D. Butt, N. Carlin, J. Lubian, C. R. Morton, J. O. Newton, A. Szanto de Toledo, Phys. Rev. C 70, 024606 (2004).
	 \bibitem{ref14} M. Dasgupta, D. J. Hinde, K. Hagino, S. B. Moraes, P. R. S. Gomes, R. M. Anjos, R. D. Butt, A. C. Berriman, N. Carlin, C. R. Morton, J. O. Newton, A. Szanto de Toledo, Phys. Rev. C 66, 041602(R) (2002).
	 \bibitem{ref15} F. Amorini, M. Cabibbo, G. Cardelaa, A. Di Pietro, P. Figuera, A. Musumarra, M. Papa, G. Pappalardo, F. Rizzo, S. Tudisco, Phys. Rev. C 58, 987 (1998). 
	 \bibitem{ref16} M. K. Sharma, P. P. Singh, D. P. Singh, A. Yadav, V. R. Sharma, I. Bala, R. Kumar, Unnati, B. P. Singh, R. Prasad, Phys. Rev. C 91, 014603 (2015).  
	 \bibitem{ref17} M. Maiti, S. Lahiri, Phys. Rev. C 79, 024611 (2009).
	 \bibitem{ref18} S. C. Srivastava, P. Richards, P. Som, G. Meinken, A. Sewatkar, T. H. Ku, Brookhaven National Laboratory Report BNL 24614 (1978).
	 \bibitem{ref19} M. C. Lagunas-Solar, M. J. Avila, N. J. Nvarro, P. C. Johnson,  Int. J. Appl. Radiat. Isot. 34, 915 (1983).
	 \bibitem{ref20}	N. G. Zaitseva,  E. Rurarz, M. Vobecky, K. H. Hwan, K. Nowak, T. Tethal, V. A. Khalkin, L. M. Popinenkova, Radiochim. Acta 56, 59 (1992). 
	 \bibitem{ref21}	D. Comar, C. Crouzel, Radiochem. Radioanal. Lett, 27, 307 (1976). 
	 \bibitem{ref22} G. Omperetto, S. M. Quim, Radiochim. Acta 27, 177 (1980).
	 \bibitem{ref23} P. J. Pao, J. L. Zhou, D. J. Silvester, S. L. Waters, Radiochem. Radioanal. Lett. 46, 21 (1981).
	 \bibitem{ref24} M. Maiti, S. Lahiri, Radiochim. Acta 99, 359-264 (2011).
	 \bibitem{ref25} M. Maiti, S. Lahiri, Radiochim Acta 103, 7–13 (2015).  
	 \bibitem{ref26} M. Maiti, Radiochim Acta 101, 437-444 (2013).
	 \bibitem{ref27} J. F.Ziegler, M. D. Ziegler, J. P. Biersack, Nucl. Instrum. Methods Phys. Res. B 268, 1818 (2010).
	 \bibitem{ref28} M. Maiti, S. Lahiri, Phys. Rev. C 81, 024603 (2010).
	 \bibitem{ref29} M. Maiti, S. Lahiri, Phys. Rev. C 84, 067601 (2011).
	 \bibitem{ref30} M. Maiti, Phys. Rev. C 84, 044615 (2011).
	 \bibitem{ref31} http://www.nndc.bnl.gov/nudat2/ (National Nuclear Data Center, Brookhaven National Laboratory).
	 \bibitem{ref32} B. Wilke, T. A. Fritz, Nucl. Instrum. Methods 138, 331 (1976).
	 \bibitem{ref33} J. Kemmer, R. Hofmann, Nucl. Instrum. Methods 176, 543 (1980).
	 \bibitem{ref34} A. Gavron, Phys. Rev. C 21, 230 (1980).
	 \bibitem{ref35} M. Blann, H. K. Vonach, Phys. Rev. C 28, 1475 (1983).
	 \bibitem{ref36} M. Blann, Lawrence Livermore National Laboratory Report No.
	 UCID 19614, 1982 (unpublished); M. Blann, paper SMR/284-1 in Proceedings of the International Centre for Theoretical Physics Workshop on Applied Nuclear Theory and Nuclear Model Calculations for Nuclear Technology Applications, Trieste, Italy, 1988 (unpublished).
	 \bibitem{ref37} M. Herman, R. Capote, B. V. Carlson, P. Oblozinsky, M. Sin, A. Trkov, H. Wienke, V. Zerkin, Nuclear Data Sheets, 108 (12), pp. 2655-2715 (2007).
	 \bibitem{ref38} R. Bass, Phys. Rev. Lett. 39, 265 (1977).
	 \bibitem{ref39} C. M. Perey, F. G. Perey, At. Data Nucl. Data Tables 17, 1
	 (1976).
	 \bibitem{ref40} M. Blann, Phys. Rev. Lett. 27, 337 (1971). 
	 \bibitem{ref41} V. F. Weisskopf, D. H. Ewing, Phys. Rev. 57, 472 (1940).
	 \bibitem{ref42} H. Lenske, H. H. Wolter, TRISTAN and ORION codes, private communication to M. Herman.
	 \bibitem{ref43} M. B. Chadwick, DDHMS code, private communication to M. Herman.
	 \bibitem{ref44} J. Raynal, Technical Report No. SMR-9/8, IAEA (unpublished).
	 \bibitem{ref45} J. Raynal, Computing as a language of physics. ICTP International Seminar Course (IAEA, ICTP, Trieste, Italy, 1972), p. 281.
	 \bibitem{ref46} C. H. Dasso, S. Landowne, Comp. Phys. Comm. 46, 187 (1987).
\end{thebibliography}
\end{document}